\documentclass[a4paper,11pt]{article}
\usepackage{pos}
\usepackage{pdftexcmds}
\makeatletter
\ifcase\pdf@shellescape
  \usepackage[frozencache]{minted}\or   % no shell escape
  \usepackage[finalizecache]{minted}\or % unrestricted shell escape
  \usepackage[frozencache]{minted}\fi   % restricted shell escape
\makeatother
\usepackage{xcolor}
\usepackage{soul}

\definecolor{ttc}{HTML}{CC2200}
\definecolor{Light}{gray}{.90}
\sethlcolor{Light}
\let\OldTexttt\texttt
\renewcommand{\texttt}[1]{{\color{ttc}\OldTexttt{#1}}}

\title{Automatic differentiation for error analysis}
\author*[a]{Alberto Ramos}
\affiliation[a]{Instituto de Física Corpuscular (IFIC),
  CSIC-Universitat de Valencia\\
  46071 - Valencia, SPAIN}

\emailAdd{alberto.ramos@ific.uv.es}

\abstract{
  We present \texttt{ADerrors.jl}, a software for linear error
  propagation and analysis of Monte Carlo data.
  Although the focus is in data analysis in Lattice QCD, where
  estimates of the observables have to be computed from Monte Carlo
  samples, the software also deals with variables with uncertainties,
  either correlated or uncorrelated. Thanks to automatic
  differentiation techniques linear error propagation is
  performed exactly, even in iterative algorithms (i.e. 
  errors in parameters of non-linear fits). 
  In this contribution we present an overview of the capabilities of
  the software, including access to uncertainties in fit parameters
  and dealing with correlated data. The software, written in julia, is
  available for download and use in~\url{https://gitlab.ift.uam-csic.es/alberto/aderrors.jl}.   
\vspace{2cm}
\begin{flushright}
IFIC/20-55
\end{flushright}
}

\FullConference{%
  Tools for High Energy Physics and Cosmology - TOOLS2020\\
  2-6 November, 2020\\
  Institut de Physique des 2 Infinis (IP2I), Lyon, France}

\begin{document}
\maketitle

\section{Data analysis in Lattice QCD}

Lattice QCD is nowadays able to provide precise results for many
quantities of phenomenological interest. 
A key ingredient in providing solid results is the analysis of lattice
QCD data. 

Lattice QCD is based on discretizing space-time in a hypercubic
grid of spacing $a$. 
The usual computation of QCD observables requires the evaluation of
path integrals. 
On the lattice these observables are computed as integrals over very
many dimensions. The biggest appeal of the method is that these
integrals can be estimated using large-scale computer simulations
and Monte Carlo methods. The usual workflow in lattice QCD analysis is
the following

\begin{description}
\item[Production of configurations] A representative ensemble of the
  discretized lattice QCD action is generated using large scale
  computer simulations. These are usually called \textbf{configurations}.
  
\item[Measurements of primary observables] Several observables of
  interest, like hadron masses ($aM_{{\rm p}, \Omega,\dots}$), decay
  constants ($af_{\pi,K,D,\dots}$), form factors, 
  etc\dots are measured in lattice units on the configurations
  generated on the previous step. 
  These are usually called \textbf{primary observables}.
  
\item[Data analysis] The observable of interest is generically a
  function of these primary observables. 
  These are called \textbf{derived observables}. 
\end{description}

The challenges in data analysis are, on one hand,
to control the systematic uncertainties associated with the
different extrapolations (to the continuum $a\to 0$, to the infinite
volume limit, etc\dots) required to produce a final answer.

On the other hand we have to correctly account
for the statistical uncertainties of the results. 
This includes a correct estimate of errors taking into
account the autocorrelation of data measured in a Monte Carlo chain. 
We also have to take into account the
correlation of observables measured on the same configurations. It is
important to note that the \emph{functions} that define the derived
observables commonly involve iterative algorithms (cf. 
section~\ref{sec:iterative-algorithms}).

In these proceedings contribution we present a freely available
analysis code, \texttt{ADerrors.jl}
(\url{https://gitlab.ift.uam-csic.es/alberto/aderrors.jl}), that aims
to simplify the data analysis. 
It combines the determination of statistical uncertainties using the
$\Gamma$-method\cite{Wolff:2003sm, Schaefer:2010hu,
  Virotta2012Critical} with automatic differentiation techniques for
error propagation~\cite{Ramos:2018vgu}. We think that some of the
techniques that are used for error propagation are potentially useful
\emph{beyond} the lattice QCD community.

\subsection{Installation}

The software is written in \texttt{Julia} (\url{https://julialang.org}). 
The package in not in the general registry. Still one can use the
package manager to install it. \texttt{ADerrors.jl} also depends on \texttt{BDIO.jl}
that is also not registered and should be installed beforehand. 
In summary the steps to have a fully functional working environment
are:
\begin{minted}[,frame=lines,linenos=false,label=Code (Julia), fontsize=\small]{julia}
julia> import Pkg
julia> Pkg.add("https://gitlab.ift.uam-csic.es/alberto/bdio.jl")
julia> Pkg.add("add https://gitlab.ift.uam-csic.es/alberto/aderrors.jl")
\end{minted}          

\section{A calculator with uncertainties}

At the center of the package \texttt{ADerrors.jl} is the data type \texttt{uwreal}. 
Generally speaking, a \texttt{uwreal} data type contains the value and
the difference sources of uncertainties of a variable. Here we list
the main concepts to work with these variables:
\begin{itemize}
\item Calling
\mintinline{julia}{var = uwreal([x,dx], "TAG")} will define
\mintinline{julia}{var} as a \texttt{uwreal} variable with central
value ${\rm x}$ and uncertainty ${\rm dx}$. 
The user provided \mintinline{julia}{"TAG"} allows to keep track of
the different contributions to the uncertainty of a variable. 
\item One can perform normal operations with \texttt{uwreal} data types, as
if they were normal real variables:
\mintinline{julia}{var2 = 2.0*sin(var)/var} will propagate errors from
\texttt{var} to \texttt{var2}.
\item Calling \mintinline{julia}{uwerr(var)} will determine the total
  uncertainty in \texttt{var}.
\item \mintinline{julia}{value(var)} and \mintinline{julia}{err(var)}
  return, respectively, the central value and total error of \texttt{var}.
\item Calling \mintinline{julia}{details(var)} will give the
  contribution to the total error of \texttt{var} from each source.
\end{itemize}

Here we show an example that shows that
\texttt{ADerrors.jl} can be used a ``calculator with uncertainties''
that automatically takes care of error propagation and correlations. 

        \begin{minted}[,frame=lines,linenos=false,label=Code (Julia), fontsize=\small]{julia}
julia> x = uwreal([12.31, 0.23], "Experiment A") # x = 12.31(23) from some experiment
12.31 (Error not available... maybe run uwerr)

julia> y = uwreal([4.22, 0.12], "Experiment B") # x = 4.22(12) from some other experiment
4.22 (Error not available... maybe run uwerr)

julia> z = x + y
16.53 (Error not available... maybe run uwerr)
julia> uwerr(z) # Determine total error in z
julia> println(z) # sqrt(0.23^2 + 0.12^2) = 0.259... (exp. A and B uncorrelated)
16.53 +/- 0.25942243542145693

julia> details(z)
16.53 +/- 0.25942243542145693
 ## Number of error sources: 2
 ## Number of MC ids       : 0
 ## Contribution to error  :               Ensemble  [%]     [MC length]
  #                                   Experiment A  78.60            -
  #                                   Experiment B  21.40            -

julia> zero_error = sin(z) - ( sin(x)*cos(y) + cos(x)*sin(y) ) 
-6.661338147750939e-16 (Error not available... maybe run uwerr)
julia> uwerr(zero_error)
julia> println(zero_error) # ADerrors keeps correlations!
-6.661338147750939e-16 +/- 1.7281005654554058e-16
        \end{minted}    

\subsection{Dealing with correlated data}

In many practical situations we have several variables with
uncertainties that are not independent. 
The covariance between these variables is given, and has to be taken
into account in order to perform error propagation correctly. 

The method \mintinline{julia}{cobs(central_values, covariance, "TAG")}
allows to import correlated data as a vector of \texttt{uwreal}. 
One can operate with the elements of this vector naturally. Let us see
an example

            \begin{minted}[,frame=lines,linenos=false,label=Code (Julia), fontsize=\small]{julia}
julia> avg = [16.26, 0.12, -0.0038]; # Central values of three quantities
julia> Mcov = [0.478071 -0.176116 0.0135305
               -0.176116 0.0696489 -0.00554431
               0.0135305 -0.00554431 0.000454180]; # Covariance matrix between the quantities

julia> p = cobs(avg, Mcov, "Correlated data") # a vector that contain the 3 quantities
3-element Array{uwreal,1}:
 16.26 (Error not available... maybe run uwerr)
 0.12 (Error not available... maybe run uwerr)
 -0.0038 (Error not available... maybe run uwerr)

julia> uwerr.(p);
julia> p
3-element Array{uwreal,1}:
 16.26 +/- 0.6914267857119798
 0.12 +/- 0.2639107803785211
 -0.0038 +/- 0.021311499243366245
julia> cov(p) # Check that we reproduce the original covariance
3×3 Array{Float64,2}:
  0.478071   -0.176116     0.0135305
 -0.176116    0.0696489   -0.00554431
  0.0135305  -0.00554431   0.00045418

julia> z = p[1] + p[2] + sin(p[3]);  # Correlations are propagated
julia> uwerr(z)
julia> z # Note that error in z is smaller than error in p[1]!
16.37620000914533 +/- 0.4603415450741195
\end{minted}

\section{Iterative algorithms}
\label{sec:iterative-algorithms}

In many practical situations our derived observables are defined by an
iterative procedure. 
This is the case of fitting parameters, that being a function of the
data that requires an iterative procedure to be determined (i.e. the
minimization of the $\chi^2$ function) make cumbersome the
determination of the derivatives required for linear error
propagation.

\texttt{ADerrors.jl} uses automatic differentiation\footnote{In fact
  \texttt{ADerrors.jl} uses the \texttt{Julia} package
  \texttt{ForwardDiff.jl}\cite{RevelsLubinPapamarkou2016}
  (\url{https://github.com/JuliaDiff/ForwardDiff.jl/}) for this
  propose.} to perform exact linear error
propagation. 
Errors can be naively propagated even in iterative algorithms. 
As a concrete example, imagine that we are interested in finding the
root of a non-linear function
\begin{equation}
  f(x) = a\cos(b\sin(x)) - x\,.
\end{equation}
If the quantities $a,b$ have uncertainties, these will
propagate into an uncertainty in the root $x^\star$. 
Since $x^\star$ is defined by
\begin{equation}
  x^\star = \text{Find the root of } f(x)\,,
\end{equation}
what error propagation needs is to find the derivative of "Find the
root of $f(x)$". 
The following program performs the job using \texttt{ADerrors.jl}
        \begin{minted}[,frame=lines,linenos=false,label=Code (Julia), fontsize=\small]{julia}
julia> a = uwreal([1.34, 0.12], "Data 01") # a = 1.34 +/- 0.12
julia> b = uwreal([1.34, 0.12], "Data 02") # b = 1.34 +/- 0.12

julia> x0 = uwreal([0.5,0.5], "Initial position") # x0 = 0.5 +/- 0.5
julia> while true # This is just newton method
           val = a*cos(b*sin(x0)) - x0
           der = -a*b*sin(b*sin(x0))*cos(x0) - 1.0
           x1  = x0 - val/der
           if (abs(value(x0) - value(x1))<1.0E-10)
               break
           else 
               x0 = x1
           end
       end

julia> uwerr(x1)
julia> details(x1)
0.7838003744331717 +/- 0.056992589665847124
 ## Number of error sources: 3
 ## Number of MC ids       : 0
 ## Contribution to error  :               Ensemble  [%]     [MC length]
  #                                        Data 02  63.25            -
  #                                        Data 01  36.75            -
  #                               Initial position   0.00            -

        \end{minted}    
Note that the \texttt{while} loop is just a straightforward
application of Newton's method. 
Automatic differentiation approaches the problem of computing the
derivative of "Find the root of $f(x)$" by just computing the 
derivative of each expression of Newton's algorithm. 
The position of the root $x^\star = 0.784(57)$, has an uncertainty
that has been propagated from the uncertainties of the original
variables $a,b$. 

Note that we also assigned an error to the initial position where the
iteration starts $x_0$. 
This uncertainty does not contribute to the final uncertainty of the
position of the root. 
This is just a manifestation of the fact that it does not matter what
we choose as starting point for our Newton method, we will end up
finding the same root. 

\subsection{Error propagation in fits}

One of the most common situations in data analysis is fitting data to
a model. This is done by finding the values of the parameters $p_i$ that
minimize the function 
\begin{equation}
\label{eq:csq}    
  \chi^2(p_i;d_a)\,,\qquad p_i\, (i=1,\dots,N_{\rm parm})\,, \quad
      d_a\, (a=1,\dots,N_{\rm data})\,.
\end{equation}
where $d_a$ are the data that carry
uncertainties (i.e. Monte Carlo data from a lattice simulation, or
experimental data). This data might be correlated and
the determination of the uncertainties in the fit parameters has to
take these correlations into account. 

The previous section shows that \texttt{ADerrors.jl} can solve this
problem by just propagating errors in each step of the minimization
routine (i.e. 
Levenberg-Marquardt, see an explicit example in~\cite{Ramos:2018vgu}). 
This is cumbersome because it would require to code this algorithm in
such a way that \texttt{uwreal} data types are accepted as input. 

Reference~\cite{Ramos:2018vgu} proposes an alternative based on the
determination of the Hessian of the $\chi^2$ function at its minima. 
This allows \texttt{ADerrors.jl} to rely on external (and efficient)
libraries to perform the minimization of the $\chi^2$, and delegate
the error propagation to the computation of the Hessian at the minima
once it is found. This Hessian can be determined exactly using
automatic differentiation techniques, allowing exact error propagation
in fit parameters.

Let us here show the main characteristics of data fitting using
\texttt{ADerrors.jl} (the complete example is available in appendix~\ref{sec:fit}). 
First we minimize the $\chi^2$ with an efficient library
\begin{minted}[,frame=lines,linenos=false,label=Code (Julia), fontsize=\small]{julia}
# Fit the data. We use LeastsquaresOptim to find the
# minima of the chisq. It expects a function of the parameters
# alone that returns the vector of residuals. 
julia> fit = optimize(xx -> lm(xx, value.(uwy)), [1.0,1.0],
                      LevenbergMarquardt(), autodiff = :forward)
julia> println("### Fit results (from LeastSquaresOptim)")
julia> println(fit)
### Fit results (from LeastSquaresOptim)
Results of Optimization Algorithm
 * Algorithm: LevenbergMarquardt
 * Minimizer: [1.19727262030669,-0.2938846629270645]
 * Sum of squares at Minimum: 19.590325
 * Iterations: 7
 * Convergence: true
 * |x - x'| < 1.0e-08: false
 * |f(x) - f(x')| / |f(x)| < 1.0e-08: true
 * |g(x)| < 1.0e-08: false
 * Function Calls: 8
 * Gradient Calls: 7
 * Multiplication Calls: 21
\end{minted}

Once the minima of the $\chi^2$ is found, \texttt{ADerrors.jl} is able
to determine the uncertainties in the fit parameters. We get the fit
parameters as a vector of \texttt{uwreal} from the routine \\
\mintinline{julia}{fit_error(csq, fitp_central_values, data)}\\
where \mintinline{julia}{csq} is the $\chi^2(p_i,d_a)$ function (as defined
in equation~(\ref{eq:csq})), \mintinline{julia}{fitp_central_values}
are the central values of the fit parameters (i.e. 
the minima of the $\chi^2$), and \mintinline{julia}{data} is the data
whose errors have to be propagated to the fit parameters.

Note that we obtain the fit parameters as an array of \texttt{uwreal}
data. 
This means that we can operate with fit parameters as if they were
normal variables, and even use them as input for other fits. 
We can also obtain the breaking down of the error in fit parameters (i.e. 
how much each original source of uncertainty contributes to the
uncertainty in fit parameters).

\begin{minted}[,frame=lines,linenos=false,label=Code (Julia), fontsize=\small]{julia}
# Propagate errors with ADerrors. Use the central values
# of fit parameters (vector fit.minimizer) as input
julia> fitp, chiexp = fit_error(csq, fit.minimizer, uwy)

# Error analysis and details of fit parameters
julia> uwerr.(fitp)
julia> println("### Error analysis from ADerrors ")
julia> println("Fit results: chi^2 / chi^2_exp: ", fit.ssr, " / ", chiexp)
julia> for i in 1:length(fitp)
           print("Fit Parameter "*string(i)*" (exact value "*string(exact[i])*": ")
           details(fitp[i])
           println(" ")
julia> end
println("### ")
### Error analysis from ADerrors 
Fit results: chi^2 / chi^2_exp: 19.59032480880994 / 18.000000000000004
Fit Parameter 1 (exact value 1.2: 1.19727262030669 +/- 0.016888186595238174
 ## Number of error sources: 20
 ## Number of MC ids       : 0
 ## Contribution to error  :               Ensemble  [%]     [MC length]
  #                     Data for point at x=0.0014  16.53            -
  #                     Data for point at x=0.0933  14.65            -
  #                     Data for point at x=0.1367  13.76            -
  #                     Data for point at x=1.8594   8.38            -
  #                     Data for point at x=0.4160   8.19            -
  #                     Data for point at x=0.5163   6.38            -
  #                     Data for point at x=1.7612   6.11            -
  #                     Data for point at x=0.5875   5.19            -
  #                     Data for point at x=0.6537   4.17            -
  #                     Data for point at x=1.6184   3.46            -
  #                     Data for point at x=0.7185   3.26            -
  #                     Data for point at x=0.7784   2.50            -
  #                     Data for point at x=0.8320   1.90            -
  #                     Data for point at x=0.8517   1.69            -
...
\end{minted}

\section{Dealing with Monte Carlo data}

Up to this point we have shown how \texttt{ADerrors.jl} works with
variables with uncertainties, even when they are correlated. 
But the most common situation in Lattice QCD analysis is that data to
be analyzed come from a Monte Carlo simulation. 
In this case one can not reduce the variable to a single \texttt{value
+/- error}, because when several observables measured on the same
ensemble are combined the correlation has to be computed from the
Monte Carlo history.

\begin{figure}
  \centering
  \includegraphics[width=\textwidth]{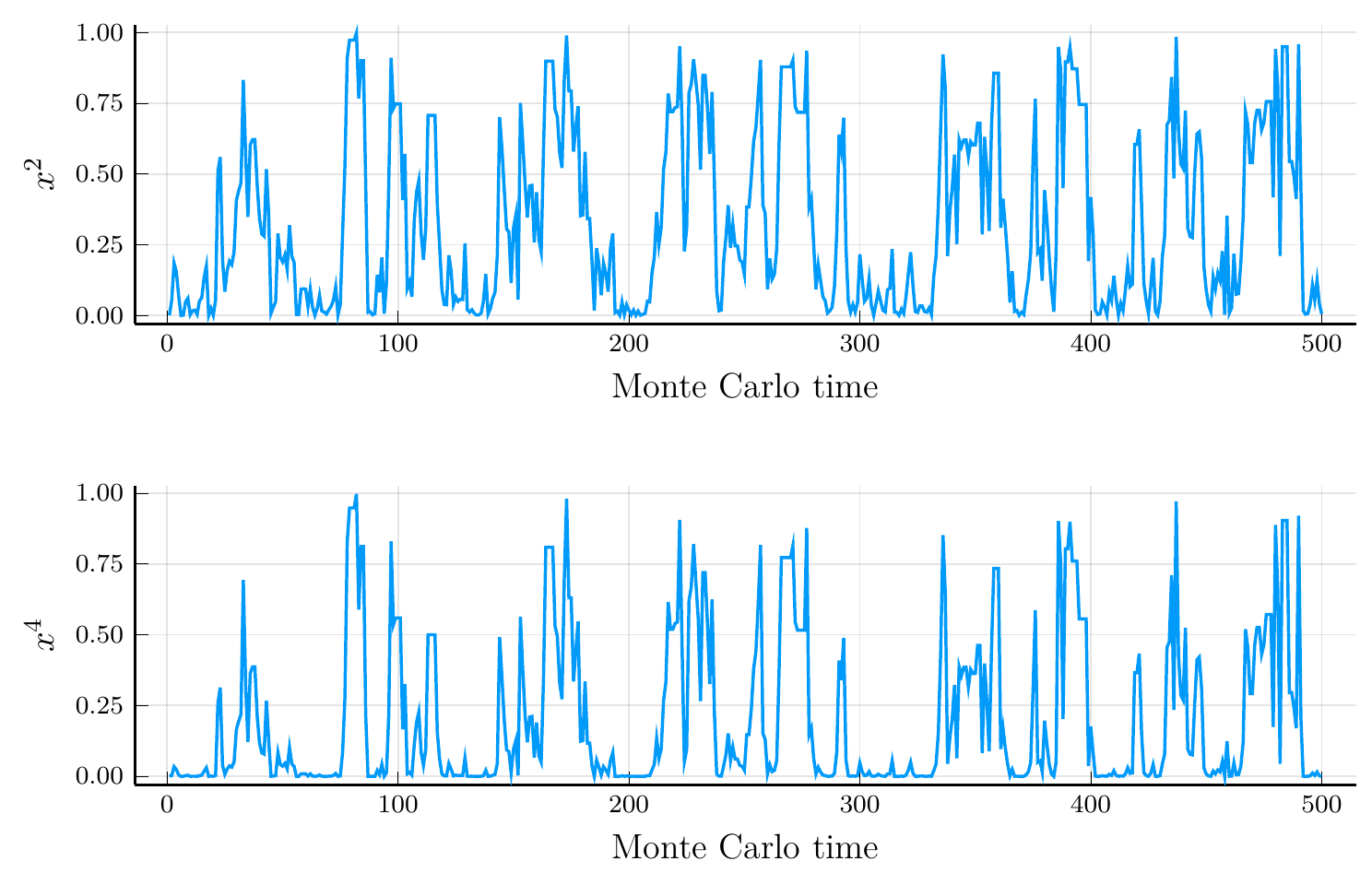}
  \caption{Monte Carlo histories of $x^2$ and $x^4$ for a random walk
    in the interval $[-1,1]$. 
  Clearly the two histories are correlated.}
  \label{fig:hist}
\end{figure}

\texttt{ADerrors.jl} deals with this case transparently:
\texttt{uwreal} variables can also contain full Monte Carlo histories. 
Let us first generate some Monte Carlo data (see figure~\ref{fig:hist})
        \begin{minted}[,frame=lines,linenos=false,label=Code (Julia), fontsize=\small]{julia}
julia> # Generate some correlated samples of the uniform[-1,1] distribution
julia> nsamp = 500
       eta  = randn(nsamp);
julia> x    = Vector{Float64}(undef, nsamp);
julia> x[1] = 0.0;
julia> for i in 2:nsamp # This is just a random walk in [-1,1]
           x[i] = x[i-1] + 0.2*eta[i]
           if abs(x[i]) > 1.0
               x[i] = x[i-1]
           end
       end
        \end{minted}    
The vector \mintinline{julia}{x[:]} contains samples of the uniform
distribution ${\rm U}[-1,1]$. We can determine the expected values,
for example, of ${\rm E}[X^2]$ or ${\rm E}[X^4]$. 
Let us show how these Monte Carlo data can be input as
\texttt{uwreal} data types and check that the variables are correlated. 
\begin{minted}[,frame=lines,linenos=false,label=Code (Julia), fontsize=\small]{julia}
julia> xp2 = uwreal(x.^2, "Random walk ensemble in [-1,1]")
0.3533602504472119 (Error not available... maybe run uwerr)

julia> xp4 = uwreal(x.^4, "Random walk ensemble in [-1,1]")
0.2197572322981959 (Error not available... maybe run uwerr)

julia> cov([xp2, xp4])
2×2 Array{Float64,2}:
 0.0007523017604391568 0.0005551868103090107
 0.0005551868103090107 0.00039631731910912886
        \end{minted}    
Crucially the ensemble tag
(\mintinline{julia}{"Random walk ensemble in [-1,1]"} in this case) is
the same for both observables. This tells \texttt{ADerrors.jl} that
the measurements of both observables have been performed in the same
ensemble and makes possible to determine the covariance of the data from the
Monte Carlo histories. 
We can determine functions of these expected values. 
Correlations are extracted from the Monte Carlo histories and
automatically taken into account. 
These correlations are crucial to obtain correct error estimates of
derived observables that depend on both variables
\mintinline{julia}{xp2, xp4}, as the following example shows
\begin{minted}[,frame=lines,linenos=false,label=Code (Julia), fontsize=\small]{julia}
julia> z = xp4/xp2^2;
julia> uwerr(z)
julia> println(z)
  1.7764683477541687 +/- 0.12751045462709376
        \end{minted}    

\section{Conclusions}

In this proceedings contribution we have presented \texttt{ADerrors.jl}. 
Using Automatic differentiation, this software package performs exact
linear error propagation even in quantities that are defined via
iterative algorithms, like fit parameters. 
\texttt{ADerrors.jl} keeps track of the origin of the error in each
variable, making possible to obtain how much each ensemble has
contributed to the uncertainty in a particular variable. 

The benefits of data analysis within this framework is that it is
\emph{robust}: 
if your analysis code determine the central values correctly, the exact nature
of automatic differentiation ensures that errors are propagated
correctly. 

Some other features of the software include
\begin{itemize}
\item Exact linear error propagation in fit parameters, integrals and
  roots of non linear functions.

\item Handles data from any number of ensembles (i.e. simulations with
  different parameters).

\item Handles correlated data, with error estimates automatically
  including the correlations. 
  
\item Support for replicas (i.e. several runs with the same simulation
  parameters).
\item Irrelgular MC measurements are handled transparently.
\end{itemize}
We refer the interested readers to the original documentation of the
package \\
\url{https://gitlab.ift.uam-csic.es/alberto/aderrors.jl}\,, \\
and the on-line tutorial \\
\url{https://ific.uv.es/~alramos/docs/ADerrors/tutorial/}\,.

\section*{Acknowledgments}

I am grateful to C. 
Pena for his help preparing the material of this talk. 
The author received support from the Generalitat Valenciana
via the genT program (CIDEGENT/2019/040).

\appendix

\section{Error propagation in fits}
\label{sec:fit}

\begin{minted}[,frame=lines,linenos=false,label=Code (Julia), fontsize=\small]{julia}
using ADerrors, Distributions, LeastSquaresOptim, Printf # hide

# First generate the data:
#  - We use 20 points
#  - The model is 1.2*exp(-x*0.3)
#  - The error in point (x,y) is 0.05/(x+y)
npt = 20
x = 2.0*rand(npt)
@. model(x) = 1.2*exp(-x*0.3)
dy = 0.05 ./ (x .+ model(x))
y = model(x) + dy .* randn(npt)

# Now the data x[:], y[:], dy[:] is input
# in ADerrors format:
uwy = Vector{uwreal}(undef, npt)
for i in 1:npt
    xfmt = @sprintf("%6.4f",x[i])
    uwy[i] = uwreal([y[i], dy[i]], "Data for point at x="*string(xfmt))
end
uwerr.(uwy)

# Show data:
println("### Data to fit ")
for i in 1:npt
    println("   dt ", x[i], " ", uwy[i])
end
println("### ")

#####
# Fit
#####
# First define the model as LeastSquaresOptim likes it.
# Also return the chisq for ADerrors.
# This routines defines the residuals and the chisq for some
# values of the x-coordinate and errors in y.
# Note: other parameters needed in the fit would have to
# be passed here. 
function fit_definitions(x, dy)

    # lmfit returns a vector. The sum of the squares is minimized
    # by LeastSquaresOptim. These are basically the residuals
    function lmfit(prm, y)
        res = Vector{eltype(prm)}(undef,length(y))
        for i in 1:length(y)
            res[i] = (y[i] - prm[1]*exp(prm[2]*x[i])) / dy[i]
        end

        return res
    end

    # This is the usual chi^2 = sum of square of residuals
    chisq(prm,data) = sum(lmfit(prm, data) .^ 2)

    return lmfit, chisq
end


# Now get our lmfit and chi^2 functions given our values of x
# and our data in uwerr format uwy
lm, csq = fit_definitions(x, err.(uwy))

# Fit the data. LeastsquaresOptim expects a function of the
# parameters alone. We fix y = value.(uwy) in the call to
# optimize. Use [1.0,1.0] as initial guess of fit parameters
fit = optimize(xx -> lm(xx, value.(uwy)), [1.0,1.0],
               LevenbergMarquardt(), autodiff = :forward)
println("### Fit results (from LeastSquaresOptim)")
println(fit)
println("### ")


# Propagate errors with ADerrors. Use the central values
# of fit parameters (vector fit.minimizer) as input
fitp, chiexp = fit_error(csq, fit.minimizer, uwy)

# Error analysis and details of fit parameters
uwerr.(fitp)
exact = [1.2, -0.3]

println("### Error analysis from ADerrors ")
println("Fit results: chi^2 / chi^2_exp: ", fit.ssr, " / ", chiexp)
for i in 1:length(fitp)
    print("Fit Parameter "*string(i)*" (exact value "*string(exact[i])*": ")
    details(fitp[i])
    println(" ")
end
println("### ")

\end{minted}

\addcontentsline{toc}{section}{References}
\bibliography{/home/alberto/docs/bib/math,/home/alberto/docs/bib/campos,/home/alberto/docs/bib/fisica,/home/alberto/docs/bib/computing}  

\end{document}